\documentclass[%
 reprint,
 amsmath,amssymb,
 aps,
 pra,
floatfix,
]{revtex4-1}

\usepackage{amsmath}
\usepackage{color}
\usepackage{physics}
\usepackage{graphicx}% Include figure files
\usepackage{dcolumn}% Align table columns on decimal point
\usepackage{bm}% bold math
\usepackage{graphicx}
\usepackage[caption=false]{subfig}
\usepackage{hyperref}% add hypertext capabilities
\begin{document}

\preprint{APS/123-QED}

\title{In-situ Raman gain between hyperfine ground states\\
				in a potassium magneto-optical trap}% Force line breaks with \\

\author{Graeme Harvie}
\author{Adam Butcher}
\author{Jon Goldwin}%
 \email{j.m.goldwin@bham.ac.uk}
\affiliation{%
 School of Physics and Astronomy, University of Birmingham,\\
 Edgbaston, Birmingham B15 2TT, United Kingdom
}%

\date{\today}% It is always \today, today,

\begin{abstract}
We study optical gain in a gas of cold $^{39}$K atoms. The gain is observed during operation of a conventional magneto-optical trap without the need for additional fields. Measurements of transmission spectra from a weak probe show that the gain is due to stimulated Raman scattering between hyperfine ground states. The experimental results are reproduced by a simplified six-level model, which also helps explain why such gain is not observed in similar experiments with rubidium or cesium.
\end{abstract}

\maketitle

%\tableofcontents

\section{\label{sec:introduction} Introduction}
It was realized early in the development of laser cooling with alkali atoms that the typical conditions in a magneto-optical trap (MOT) support steady-state optical gain \cite{Tabosa1991, Grison1991}. Multiple mechanisms have been observed in both cesium and rubidium MOTs. For probe frequencies near the cooling laser frequency, a narrow dispersively shaped spectral feature arises due to Raman scattering between $m_F$ states of a single hyperfine level $F$, the degeneracy of which is lifted by polarization-dependent light shifts \cite{Grison1991, Hilico1992, Guerin2008}. Under certain conditions, in-situ four-wave mixing and recoil-induced resonances can also be observed \cite{Guo1992, Brzozowski2005, Guerin2008}. At a probe frequency near twice the detuning of the cooling laser from the cycling transition, a broader feature is observed. This feature is due to Mollow gain between dressed states \cite{Mollow1972, Yu1977, Mitsunaga1996, Guerin2008, Sawant2017}. It is particularly notable that these gain mechanisms do not require additional lasers beyond those used for cooling and trapping, suggesting a relatively simple setup for continuous-wave amplification or lasing with cold atoms.

\begin{figure}
    \centering
    \includegraphics[width=0.8\columnwidth]{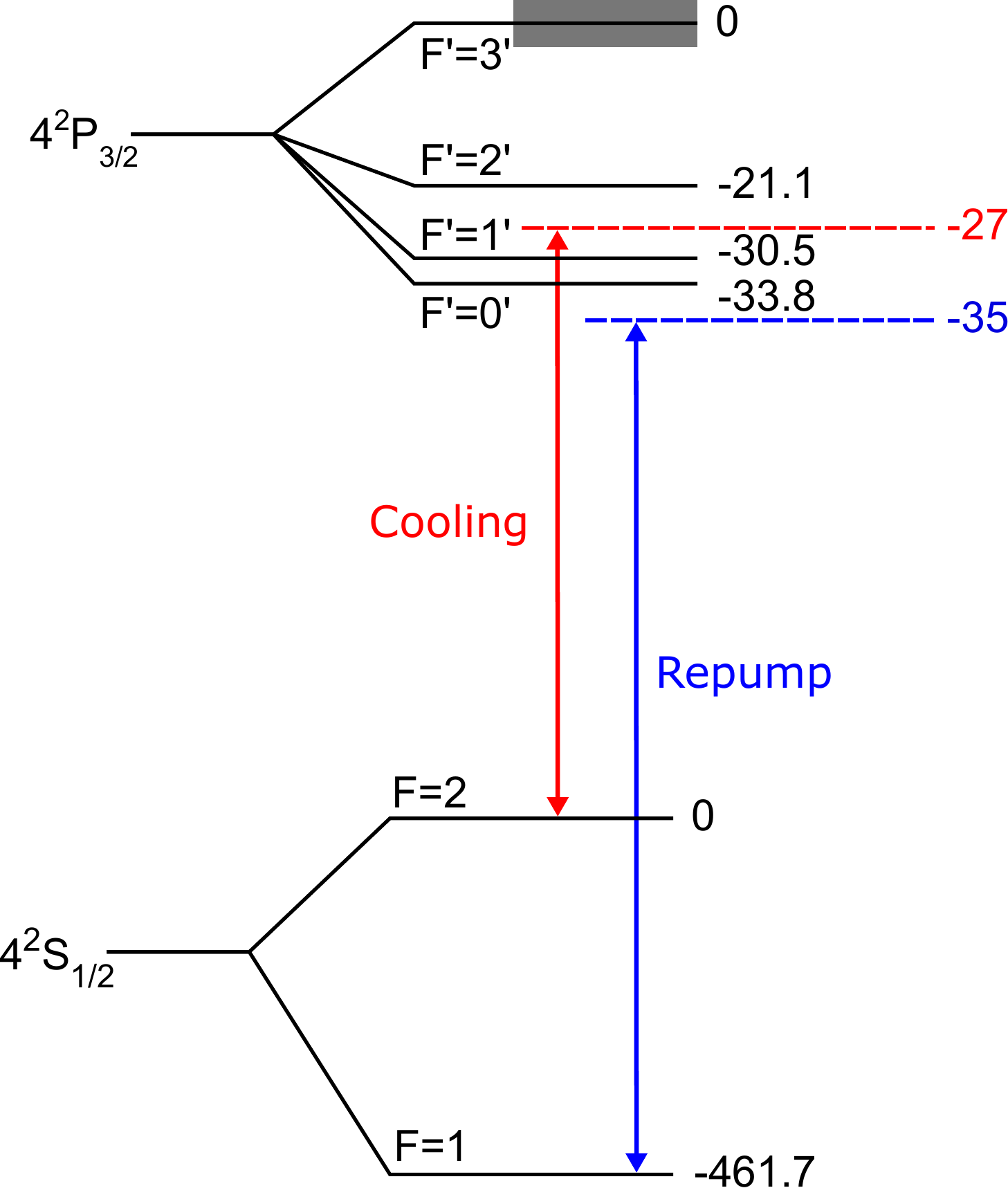}
    \caption{Energy level diagram for the D2 lines of $^{39}$K at 767~nm (the excited- and ground-state manifolds and their separation are not to scale with each other). The MOT cooling light is represented in red and the repump in blue. Frequencies are given in units of MHz, and the indicated detunings are the default values for making a MOT with a large atom number in our experiment. The natural linewidth is $6.0$~MHz full-width at half-maximum (FWHM), denoted by the grey bar around the $F'=3'$ level.}
    \label{fig:energy_level_diagram}
\end{figure}

Recently we observed steady-state lasing with cold $^{39}$K atoms in a ring cavity \cite{Megyeri2018}. To our knowledge this was the first demonstration of gain in an operating potassium MOT. Compared to cesium or rubidium, $^{39}$K has relatively small excited-state hyperfine splittings, being on the order of the natural linewidth; a simplified energy level schematic is shown in Fig.~\ref{fig:energy_level_diagram}. This means a potassium MOT operates in a significantly different way, in that light scattering is dominated by a non-cycling transition --- the primary cooling laser in our experiment preferentially drives the $\ket{F=2}\to\ket{F'=2'}$ transition (primes denote excited states), which decays by spontaneous emission to the $\ket{1}$ ground state with $50\%$ probability. Because of this, $^{39}$K MOTs require more repump light than their rubidium or cesium counterparts, and the fractional population in the lower ground state can be much larger.

In our experiment, we observe no evidence of the Raman gain or recoil-induced resonances between nearly-degenerate Zeeman states which are familiar from cesium and rubidium. We do observe a broader gain feature which we previously associated with Mollow gain on the $\ket{2}\to\ket{2'}$ transition. However, subsequent measurements, presented in detail here, suggest that the gain is in fact due to Raman scattering between the $\ket{1}$ and $\ket{2}$ ground states. The rest of this paper is organized as follows. In Section \ref{sec:spectra} we describe the experimental apparatus, and present measurements of transmission spectra for varying detunings of the MOT cooling and repump light. In Section \ref{sec:num_mod} we present a six-level model for hyperfine Raman gain which is in good agreement with the measured spectra. Motivated by this model, we investigate the role of the ground state populations in Section \ref{sec:pops}. Finally in Section \ref{sec:conc} we return to a comparison with rubidium and cesium and give an outlook for future work.

\section{\label{sec:spectra} Transmission Spectra}
Our experiment is described in detail in \cite{Culver2016, rob_thesis, andreas_thesis, Megyeri2018}, so we only give a brief overview here. Potassium atoms released from a heated ampule are cooled in a 2D-MOT, and continuously loaded into a 3D-MOT with a quasi-resonant pushing laser. The cooling and repump lasers are near resonance with the groups of $\ket{2}\to\ket{F'}$ and $\ket{1}\to\ket{F'}$ transitions, respectively (recall Fig.~\ref{fig:energy_level_diagram}). As mentioned above, the designations of cooling and repump light do not accurately apply to a $^{39}$K MOT, but we use them here for consistency with other work. The repump light is derived from the same laser as the cooling light using an acousto-optic modulator (AOM), and both beams are amplified with a single tapered amplifier. The amplified light is coupled into polarization-maintaining single-mode optical fibers for spatial mode cleaning and delivered to the atoms in a conventional three-beam, retro-reflected geometry; the cooling and repump beams have similar peak intensities of $\approx 14$~mW/cm$^2$ at the position of the atoms. A quadrupole magnetic field with a gradient of 11~G/cm is on during all experiments. Under typical conditions the temperature of the cloud is around 1~mK.
   
A linearly polarized probe beam derived from an independent laser is used to measure the transmission spectrum of the running MOT. The probe passes through the center of the MOT, and is focused such that the beam waist of $\approx 100~\mu$m is much smaller than the MOT radius (typically $800~\mu$m), ensuring the probe traverses the maximum column density of atoms. The probe power is $\approx 1 \mu$W, which does not noticeably perturb the running MOT. The corresponding probe intensity at the atoms is $\approx 1$~mW/cm$^2$, which is just below the saturation intensity of $1.75$~mW/cm$^2$ calculated for the $\ket{2}\to\ket{3'}$ cycling transition. The probe propagates nearly parallel to one pair of MOT beams, then passes through a $4.1$~mm aperture to filter out stray MOT light, before being detected on an avalanche photodiode (Thorlabs APD110A/M). The probe laser is free running, with its frequency scanned over a range of at least 1~GHz, allowing all of the D2 features to be observed in a single scan. To calibrate the frequency sweep rate, a fraction of the probe light is incident on a high finesse optical cavity. Sidebands can be added to the probe at $17.3$~MHz by modulating the laser current, and the resulting sidebands in the cavity transmission spectrum can then be used as a frequency reference for the probe scan. The modulation sidebands are turned off during MOT transmission measurements.
    
\begin{figure}
    \centering
    \includegraphics[width=\columnwidth]{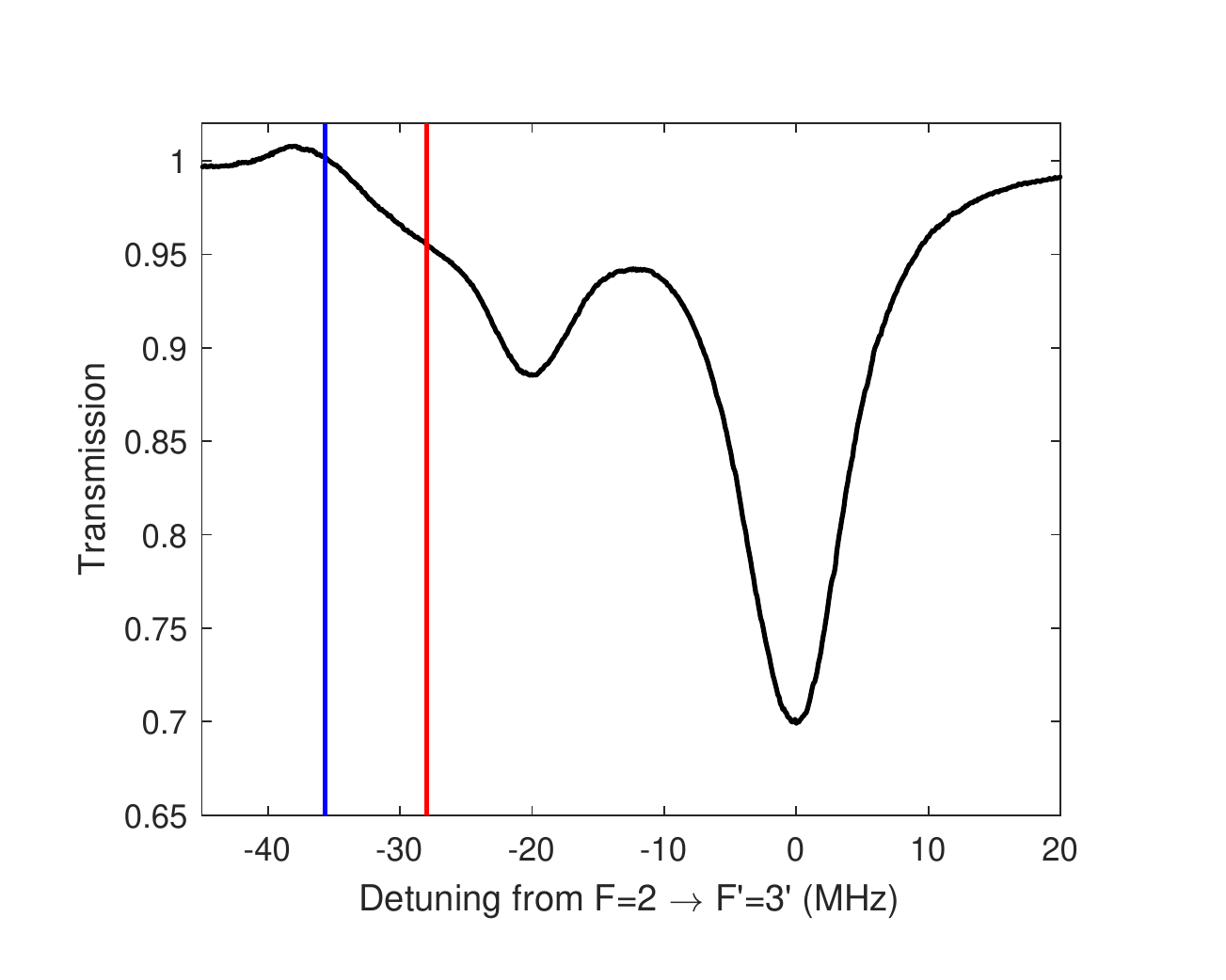}
    \caption{Fractional probe transmission spectrum obtained using the default MOT beam detunings shown in Fig.~\ref{fig:default_spec}. The cooling laser frequency and the expected hyperfine Raman gain frequency are indicated by the red and blue lines, respectively.}
    \label{fig:default_spec}
\end{figure}
    
An example spectrum obtained using our default MOT detunings is shown in Fig.~\ref{fig:default_spec}. To obtain a normalized transmission spectrum, the photodiode signal is first averaged 32 times on a digital storage oscilloscope; a linear fit to the wings of the spectrum accounts for the intensity change of the laser during the frequency scan, and the probe beam is then blocked in order to make a measurement of the background. We observe two large absorption features corresponding to the $\ket{2}\to\ket{3'}$ and $\ket{2}\to\ket{2'}$ transitions, a smaller $\ket{2}\to\ket{1'}$ absorption feature, and a single gain feature to the red. The gain feature is qualitatively similar to what one would expect for Mollow gain on the $\ket{2}\to\ket{2'}$ transition, in the sense that the linewidth is on the order of the natural linewidth, and the center frequency is approximately double the detuning of the cooling beam from the $\ket{2}\to\ket{2'}$ absorption resonance (indicated by the red line). However the gain feature is also close to the two-photon Raman transition between the ground states (blue line). This suggests an alternative explanation --- the gain could be caused by a stimulated Raman transition from $\ket{1}\to\ket{2}$, with the MOT repump light acting as the Raman pump. The small offset between the expected Raman transition frequency and the gain feature is comparable to the calculated Stark shifts of the ground states from the MOT beams. 

In order to differentiate between Mollow and hyperfine Raman gain, further measurements were taken with varying detunings for the MOT cooling and repump beams. The two mechanisms are expected to exhibit complementary behavior with respect to the laser frequencies --- neglecting relatively small light shifts, the frequency of the Raman gain peak should depend only on the MOT repump laser frequency, while Mollow gain should depend only on the frequency of the cooling light.

\begin{figure*}
    \centering
    \subfloat{%
        \includegraphics[clip,width=\columnwidth]{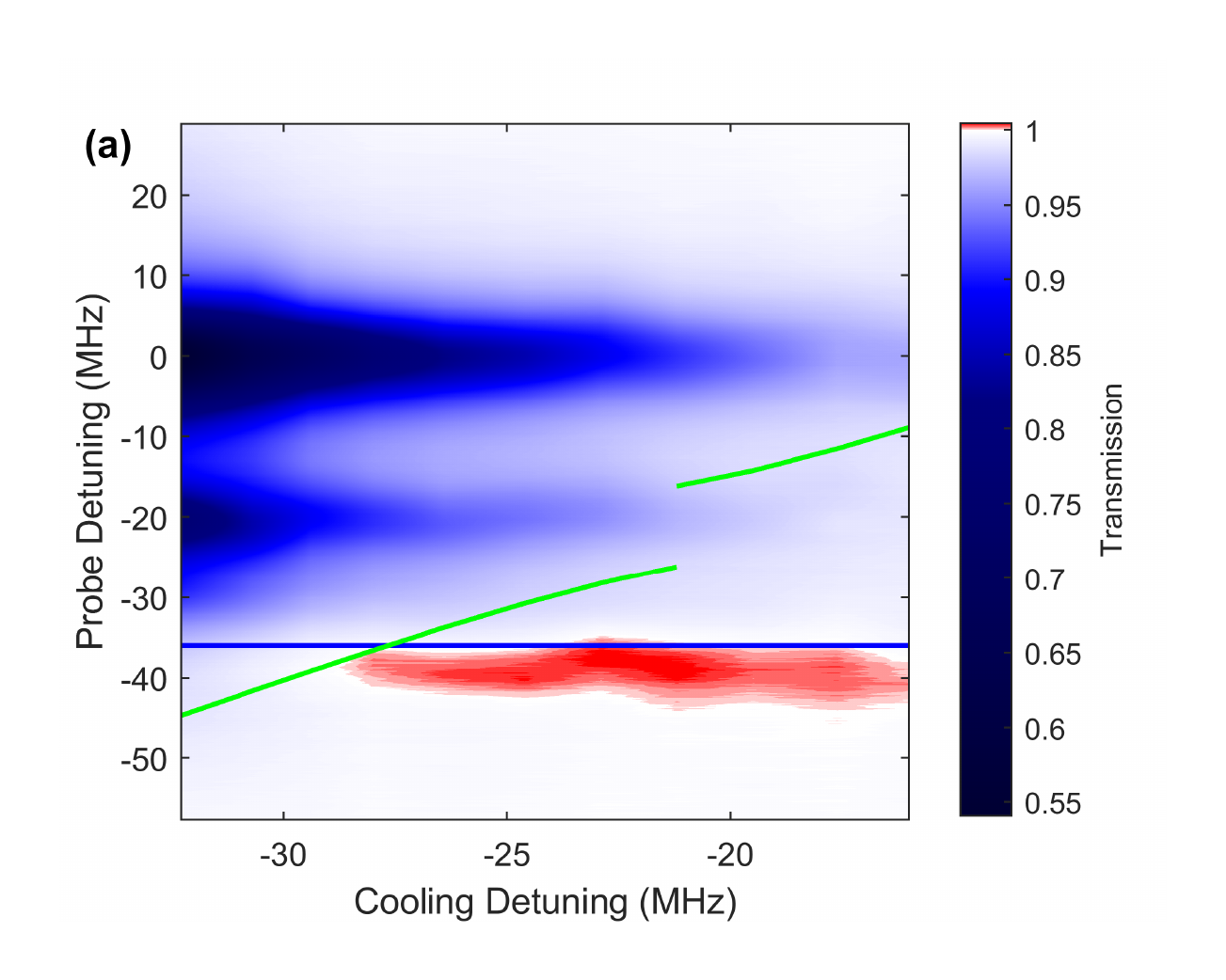} %\label{fig:cool_vs_gain_contour}%
    }
    \subfloat{%
        \includegraphics[clip,width=\columnwidth]{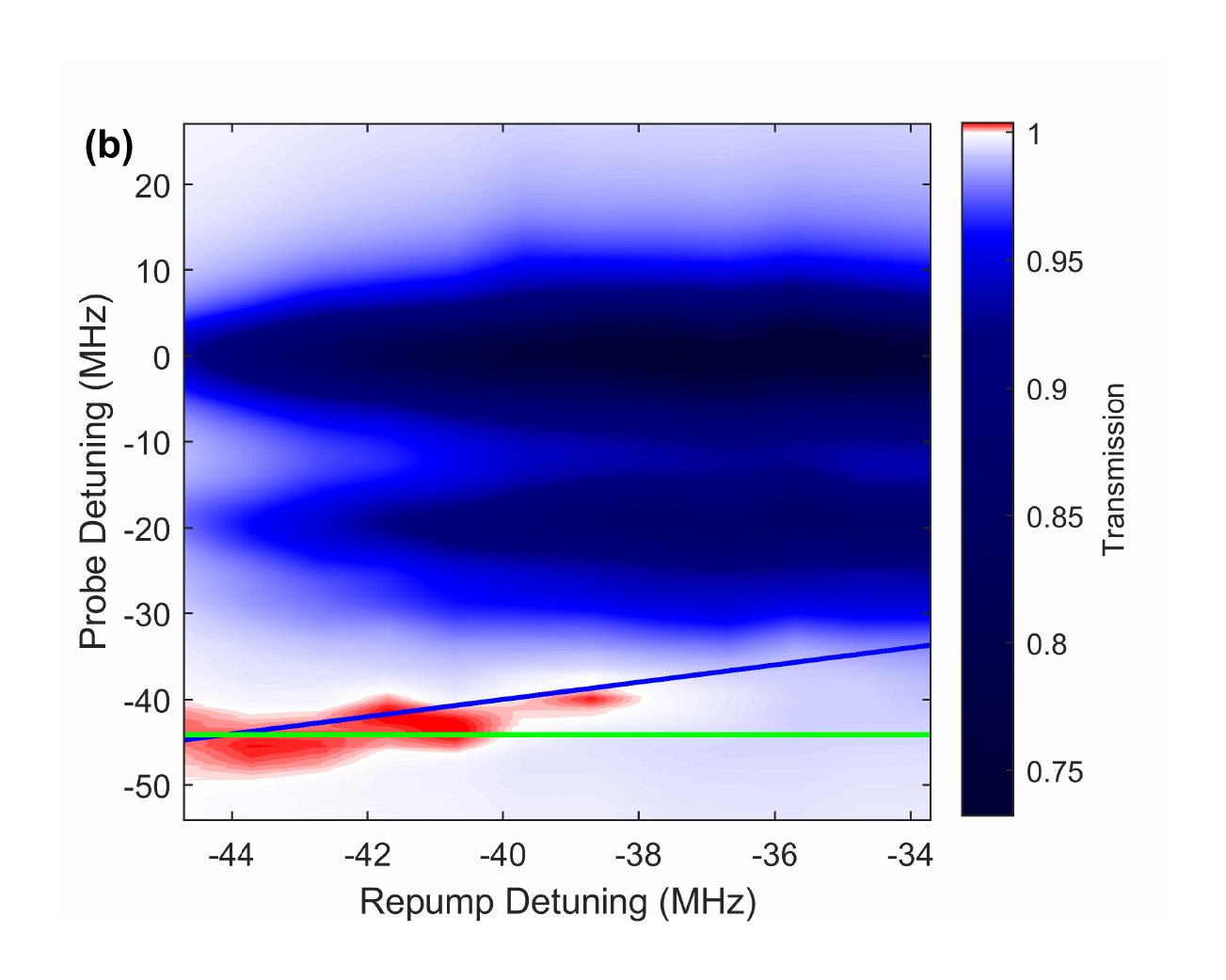}
        %\label{fig:rp_vs_gain_contour}%
    }
    \caption{Absorption and gain for varying MOT beam frequencies. (a) Measured probe transmission as the cooling frequency is changed. Absorption is shown in blue and gain is shown in red. Note the difference in scale between red and blue in the colorbars. The green curve shows the expected resonance frequency for Mollow gain on $\ket{2}\to\ket{2'}$ and the blue line the $\ket{1}\to\ket{2}$ Raman resonance frequency in the absence of AC Stark shifts. (b) Measured probe transmission as the repump laser frequency is changed. \label{fig:contours}}
\end{figure*}
    
Figure \ref{fig:contours} shows that the gain resonance frequency is primarily determined by the repump laser frequency, implying Raman gain. This is most evident in Fig.~\ref{fig:contours}(a), which shows no observable dependence on the cooling laser frequency, even when the cooling frequency crosses the $\ket{2} \rightarrow \ket{2'}$ resonance. This is in sharp contrast to the avoided crossing one would expect from Mollow gain (green curve) \cite{Mitsunaga1996}. For comparison, Fig.~\ref{fig:contours}(b) shows that the gain frequency closely follows the two-photon Raman resonance as the repump detuning is varied; Mollow gain would lead to a similar resonance frequency for these parameters, but one which would not depend on the repump detuning. Note that the cooling detuning was increased to $-31$~MHz for this measurement. This reduces the rate of optical pumping into the $\ket{1}$ state, which allows a wider range of repump detunings to be explored while maintaining a reasonable number of atoms in the MOT. 

The amplitudes of the gain and absorption features also depend on the cooling and repump frequencies. This is due in part to changes in the total number of atoms in the MOT. Interestingly, we observe that the gain amplitude is higher when the absorption amplitude is lower, and vice versa. The theoretical model presented in the next section shows that this is expected for Raman gain between hyperfine ground states. This is because the amount of absorption is proportional to the population in the $\ket{2}$ state, whereas the gain amplitude is proportional to the population difference between the $\ket{1}$ and $\ket{2}$ states \cite{Vrijsen2011}.

It is worth noting that Fig.~\ref{fig:contours}(a) appears to contradict our previous measurements, where the gain resonance moved with the cooling laser frequency (denoted as `pump' detuning in Fig.~S1(a) of the supporting material for \cite{Megyeri2018}). In that work the repump frequency was also changing, with a fixed frequency offset relative to the cooling beam, set by the AOM used to derive the repump light. Based on previous work with cesium and rubidium MOTs, we did not foresee that the repump light itself could act as the gain pump. In the work presented here we have been careful to vary the frequency of only one field at a time --- when the cooling laser frequency is changed in Fig.~\ref{fig:contours}(a), the AOM frequency is counter-tuned to maintain a fixed repump frequency; in Fig.~\ref{fig:contours}(b), only the repump frequency is varied, by keeping the cooling laser frequency fixed and tuning the AOM. These new measurements show unambiguously that the gain resonance frequency is determined by the repump frequency in our experiment.
    
\section{\label{sec:num_mod} Theoretical Model}

In order to better understand the observed spectra, we reduce the system to a six-level model consisting of the hyperfine levels $F=1,2$ and $F'=0'$--$3'$, but neglecting the nearly degenerate Zeeman sub-states. A Lindblad-type master equation is used to solve for the density matrix $\rho$,
\begin{equation}
    \dot{\rho} = \frac{1}{i\hbar}[H,\rho] + \sum_n\frac{\gamma_n}{2}(2a_n\rho a_n^{\dag}-a_n^{\dag}a_n\rho-\rho a_n^{\dag}a_n)\; ,
    \label{eq:master_equation}
\end{equation}
where $H$ is the Hamiltonian. Each $n$ represents a channel for incoherent evolution of the system; $a_n$ is the corresponding collapse operator, and $\gamma_n$ the associated rate constant. The steady state of the system is found by solving for $\dot{\rho}=0$. 

In the dipole approximation and interaction picture,
\begin{align}
    H &= \hbar\delta\ket{2}\!\!\bra{2} - \sum_{F'}\hbar \Delta_{F'} \ket{F'}\!\!\bra{F'} \nonumber\\ 
    &- \sum_{F'} \frac{\hbar\Omega_{1F'}}{2}\Big(\ket{1}\!\!\bra{F'} + \ket{F'}\!\!\bra{1}\Big) \label{eq:sl_mod_ham} \\
    &- \sum_{F'} \frac{\hbar\Omega_{2F'}}{2}\Big(\ket{2}\!\!\bra{F'} + \ket{F'}\!\!\bra{2}\Big)\quad. \nonumber
\end{align}
Here $\delta$ is the two-photon Raman detuning between the probe and the repump, and $\Delta_{F'}$ is the repump detuning from the unperturbed $\ket{1}\to\ket{F'}$ transition. The Rabi frequency $\Omega_{1F'}$ describes the repump light driving the $\ket{1}\to\ket{F'}$ transition, while $\Omega_{2F'}$ describes the weak probe acting on $\ket{2}\to\ket{F'}$. In calculating the Rabi frequencies, we use effective Clebsch-Gordan coefficients $C_{FF'}$ for $\pi$ transitions averaged over all possible Zeeman states as in \cite{Megyeri2018}. 

The cooling light is included in Eq.(\ref{eq:master_equation}) as an incoherent pumping term from the $\ket{2}$ state to the $\ket{1}$ state with effective collapse operator $\ket{1}\!\!\bra{2}$ and rate $w$. This replicates the effect of the cooling beam in repopulating the $\ket{1}$ state, without giving rise to additional features due to coherent scattering within the numerous $\Lambda$ systems involved \cite{Birnbaum2006}. A similar approach was adopted in Refs.~\cite{Meiser2009, Vrijsen2011}. An added benefit of this is that it allows a straightforward calculation of the transmission spectrum by including a weak probe term in the Hamiltonian, without causing beating with the cooling light. Spontaneous emission is represented by collapse operators $\ket{F}\!\!\bra{F'}$ with rates $\Gamma\,b_{FF'}$, where $\Gamma=2\pi\times 6.0$~MHz is the natural linewidth (FWHM) and $b_{FF'}$ are branching ratios. Dephasing between ground states is described by a collapse operator for the population difference $(\,\ket{1}\!\!\bra{1}-\ket{2}\!\!\bra{2}\,)$ with rate $\gamma_{12}\,$. The steady-state solution for $\rho$ is solved numerically using the QuTiP software package \cite{JOHANSSON2013}.

The linear probe susceptibility for the $\ket{2}\to\ket{F'}$ transition is proportional to $\rho_{2F'}$, with the imaginary part giving the absorption or gain. The probe transmission for an optically thin MOT is finally given by,
\begin{equation}
     T =  1-A\sum_{F'} C_{2F'} \Im\{\rho_{2F'}\}\quad, 
 \end{equation}
 where $A$ is an overall scaling factor used to match the peak absorption to the measured value. This is ultimately determined by the column density of the MOT.
 
 \begin{figure}
     \centering
     \includegraphics[width=\columnwidth]{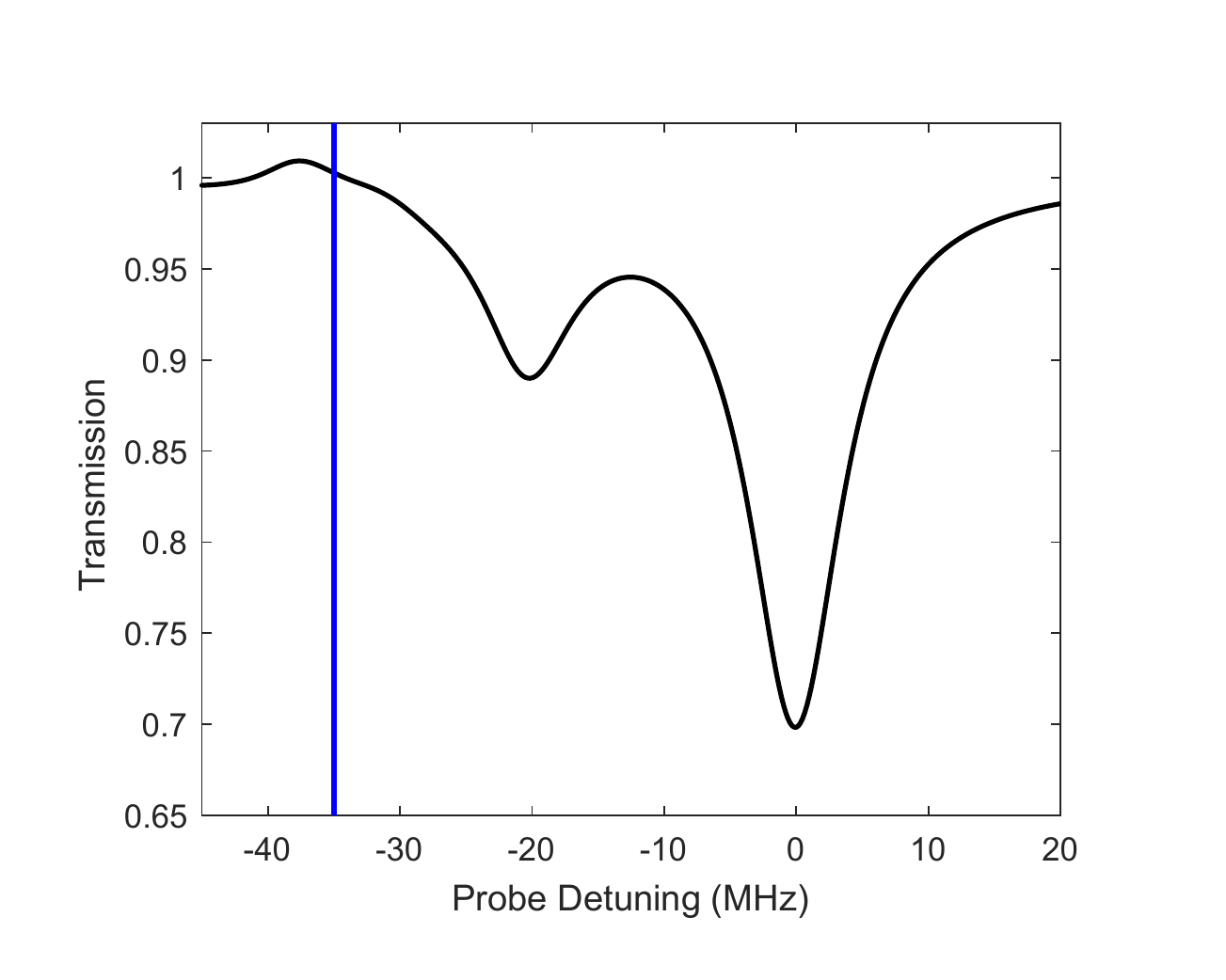}
     \caption{Calculated transmission spectrum, with the repump detuning shown in Fig.~\ref{fig:energy_level_diagram}. The phenomenological parameters used were $w=2\pi\times 1.0$~MHz and $\gamma_{12}=2\pi\times 1.5$~MHz. The rest of the parameters matched the experimental conditions.}
     \label{fig:num_spec}
 \end{figure} 
 
A calculated probe spectrum is shown in Fig.~\ref{fig:num_spec}. The parameters $A$, $w$, and $\gamma_{12}$ were set to match the measured spectrum in Fig.~\ref{fig:default_spec}. Increasing $A$ increases the amplitude of all features by a common factor; the amplitude of the gain peak increases with $w$ and decreases with $\gamma_{12}$, while the width of the gain peak increases with both. We obtain a good match to the data for $w=2\pi\times 1.0$~MHz and $\gamma_{12}=2\pi\times 1.5$~MHz. The optical pumping rate $w$ can be estimated from the optical Bloch equations, which predict $w=2\pi\times 0.6\,$-$0.8$~MHz for our conditions, depending on the fraction of the total MOT light intensity on the cooling transitions, which can vary due to competition for gain in the tapered amplifier. Dephasing at a rate $\gamma_{12}$ leads to broadening of the simulated lines, such that the half-widths at half-maximum increase from approximately $\tfrac{1}{2}(\Gamma+w)\to\tfrac{1}{2}(\Gamma+w+\gamma_{12})$ for absorption and $\tfrac{1}{2}w\to\tfrac{1}{2}w+2\gamma_{12}$ for gain. To compare the simulations to the data, we assume Voigt lineshapes for the measured features, with Lorentzian widths corresponding to the left-hand sides of these expressions, broadened by a common Gaussian width $\eta$ (HWHM). For the data presented here, $\eta$ is dominated by relative frequency fluctuations between the MOT laser and the free-running probe laser. We measure a Gaussian beat note with a half-width of $2\pi\times 1.8$~MHz. To estimate the effects of Zeeman shifts in the spatially varying magnetic field, we first calculate the root-mean-squared value of $(g_2m_2-g_1m_1)$ over the 40 possible Raman transitions between $\ket{1}$ and $\ket{2}$, weighted by the appropriate transition strengths (here the $g_F$ are Land\'{e} $g$-factors and the $m_F$ are magnetic quantum numbers). This gives an effective scale factor of $(25/24)^{1/2}\mu_B/\hbar$. Then averaging the quadrupole gradient magnetic field over the Gaussian density of atoms along the probe beam direction, we obtain a half-width of $2\pi\times 0.6$~MHz. Finally, we expect a contribution due to Doppler broadening. In the case of Raman scattering, this depends on the angle between pump and probe beams, which is approximately 0, 90, or 180 degrees for each of the six MOT beams. The resulting Doppler widths are in the range of 0 to 2 times the single-photon value of $2\pi\times 0.7$~MHz. Combining all of these effects, we estimate $\eta\approx 2\pi\times 2.1$~MHz. Together with the known value of $\Gamma$ and the simulated value of $w$, the corresponding Voigt linewidths for absorption and gain imply $\gamma_{12}=2\pi\times 0.8\,$-$2.3$~MHz.

\begin{figure}
    \centering
    \includegraphics[width=\columnwidth]{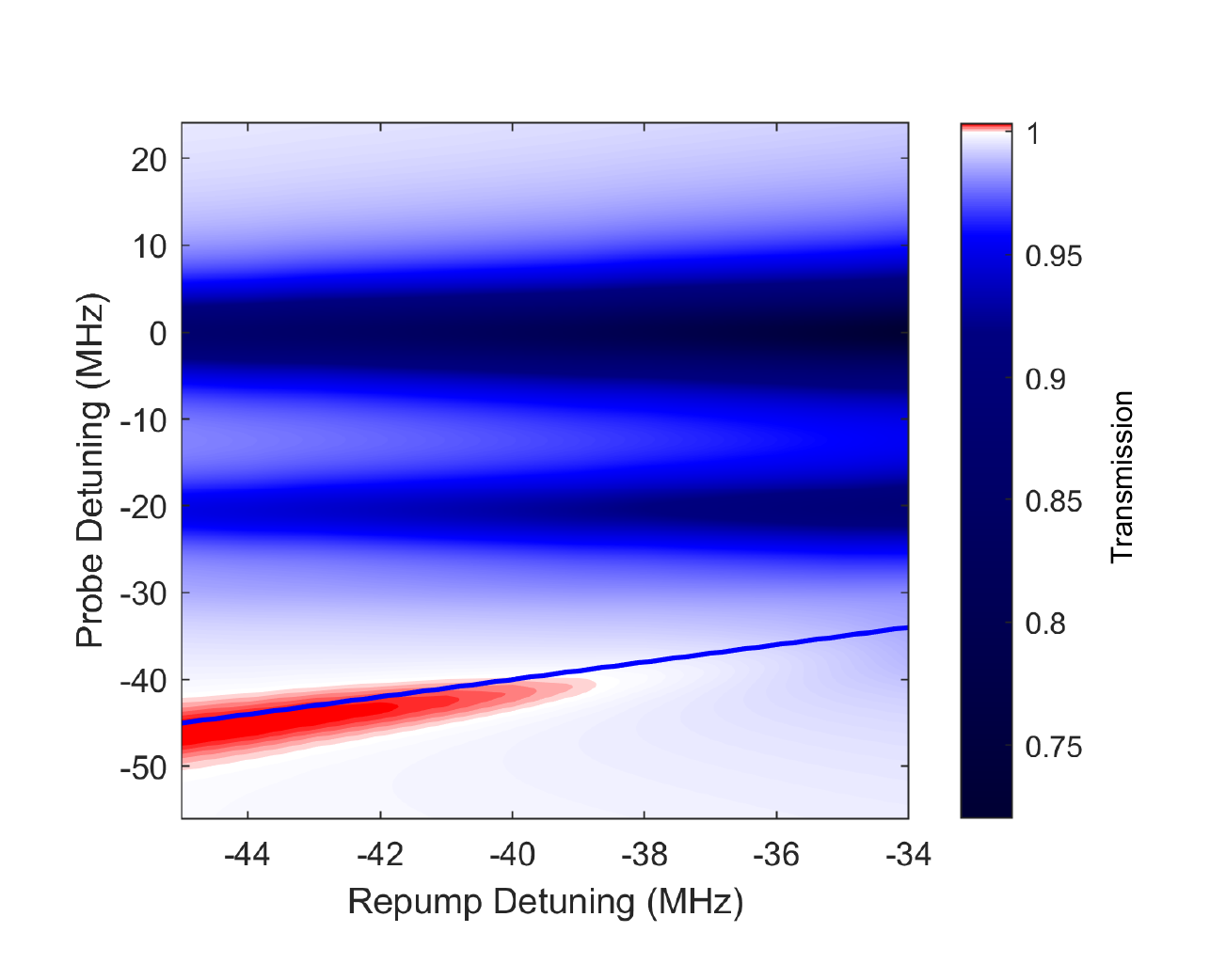}
    \caption{Calculated probe transmission for varying repump detuning. Parameters were chosen to match the experimental conditions of Fig.~\ref{fig:contours}(b), but with fixed atom number. The optical pumping rate was $w=2\pi\times 0.5$~MHz and the dephasing rate was $\gamma_{12}=2\pi\times 1.5$~MHz.}
    \label{fig:rp_vs_gain_contour_num}
\end{figure}

To further compare the theoretical model to the experiment, a contour plot similar to Fig.~\ref{fig:contours}(b) was made. This is shown in Fig.~\ref{fig:rp_vs_gain_contour_num}, with all parameters, including the total atom number, fixed. This can also be seen to match the experimental results well. In particular, the increase in peak gain and reduction in absorption with increasing repump detuning is replicated in the numerical model. Note that the value of $w$ used in this simulation is lower than in Fig.~\ref{fig:num_spec}, reflecting the fact that the cooling beam was further detuned in the corresponding measurement.
 
Despite some apparent similarities, the model presented here is qualitatively different than the four-level model used in \cite{Megyeri2018} under the assumption of Mollow gain on the $\ket{2}\to\ket{2'}$ transition. Specifically, the present model only describes Raman gain pumped by the repump light, whereas the four-level model only describes Mollow gain pumped by the cooling light. The fact that both models predict gain for our experiment suggests that, in principle, both gain mechanisms could coexist in our experiment. However the four-level Mollow model does not include the $\ket{1}$ ground state, and therefore cannot account for optical pumping out of $\ket{2}$. This pumping suppresses Mollow gain and increases Raman gain, as described in more detail in the next section. We have found that all of our experimental results are well described by pure Raman gain induced by the repump light for the parameter regimes we have studied.

\section{\label{sec:pops}Ground state populations}

\begin{figure}[ht]
    \centering
    \includegraphics[width=\columnwidth]{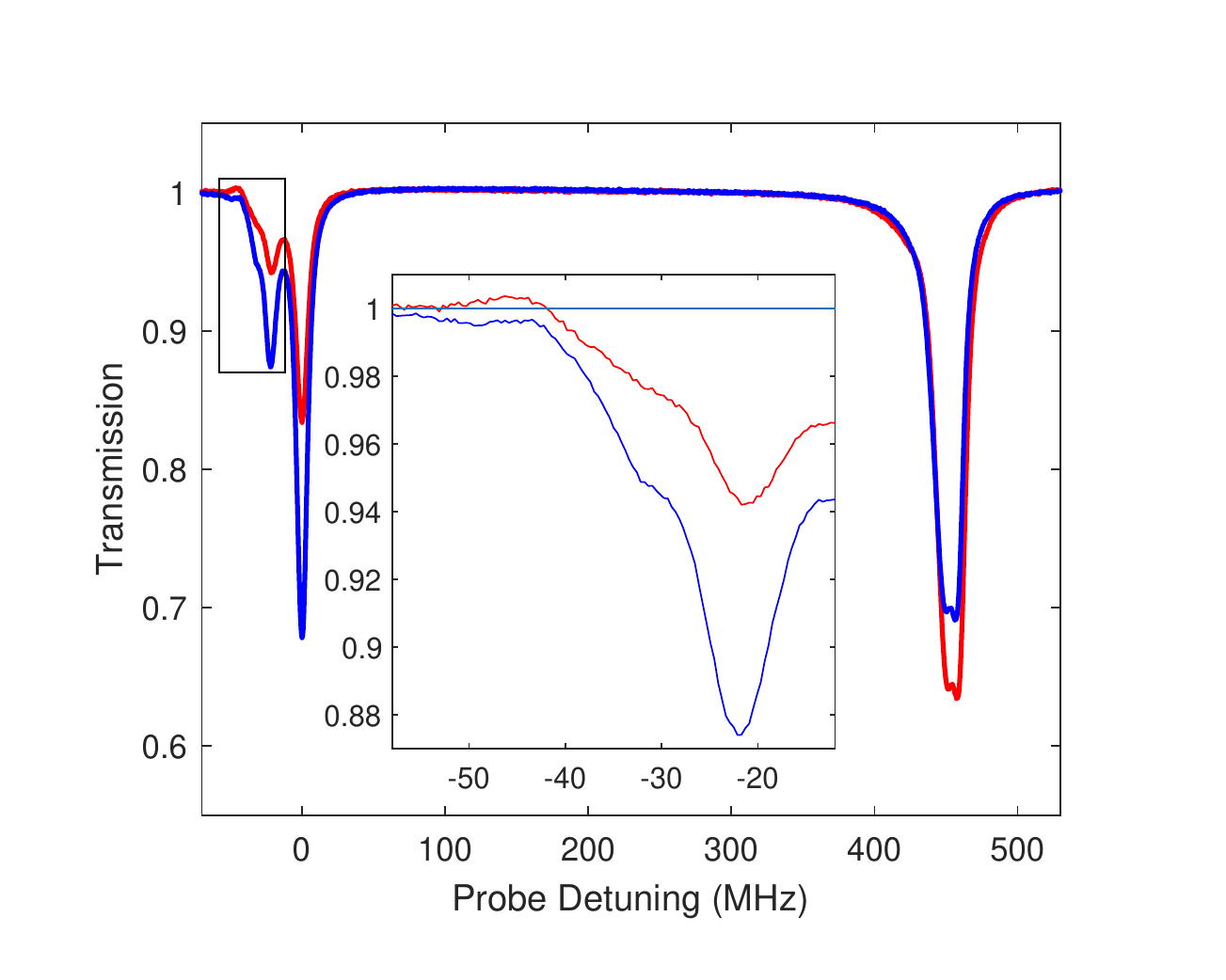}
    \caption{Measured transmission spectra used to obtain the ground state populations. The red and blue curves correspond to repump detunings of $-34.7$ and $-30.7$~MHz, respectively. The features around zero detuning are the gain and $\ket{2}\to\ket{F'}$ absorption resonances, and the feature around 460~MHz comprises the overlapping $\ket{1}\to\ket{F'}$ transitions. The inset shows a detail of the region between the gain and $\ket{2}\to\ket{3'}$ absorption resonances.}
    \label{fig:rep_probe_spectra}
\end{figure}

Hyperfine Raman gain is proportional to the population difference between the $\ket{1}$ and $\ket{2}$ ground states \cite{Vrijsen2011}. These populations should therefore be measured, in order to ensure they are consistent with the theoretical model. The populations can be determined from transmission spectra spanning all of the D2 transitions. Example spectra are shown in Fig.~\ref{fig:rep_probe_spectra} for the default cooling frequency and two different values of the repump frequency.

\begin{figure}
    \centering
    \includegraphics[width=\columnwidth]{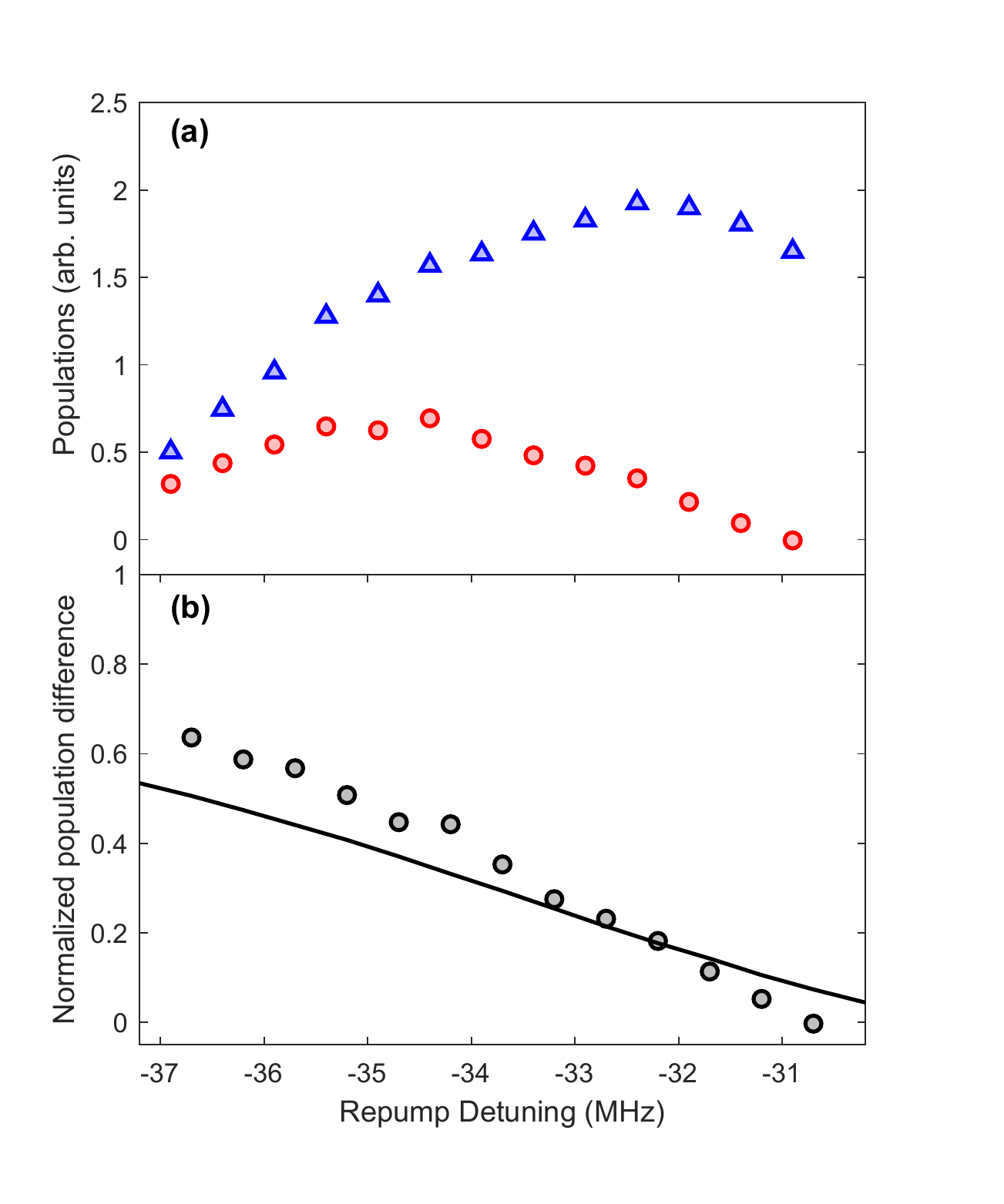}
    \caption{Ground state populations versus repump detuning. (a) The experimentally determined population difference $n_{\rm diff}=n_1-n_2$ (red circles) and sum $n_{\rm sum}=n_1+n_2$ (blue triangles), where $n_i$ is the column density of atoms in ground state $\ket{i}$. (b) The normalized population difference $n_{\rm diff}/n_{\rm sum}$, from experiment (points) and theory (line).}
    \label{fig:rep_pops}
\end{figure}

The dominant $\ket{2}\to\ket{3'}$ and combined $\ket{1}\to\ket{F'}$ absorption features have similar optical depths when the populations are equal (relatively $0.47$ and $0.45$, respectively). These features are therefore used for estimating the populations by normalizing the measured optical depth to the square of the relevant Clebsch-Gordan coefficient. The populations determined in this way are shown in Fig.~\ref{fig:rep_pops}(a) for varying detuning of the repump light. The data show that the population difference goes to zero at small repump detunings, which is the reason for the reduction in peak gain observed in Fig.~\ref{fig:rp_vs_gain_contour_num}. For increasing repump detuning, the population difference reaches a maximum and then drops off due to the reduction in total atom number in the MOT.  The theoretical model does not include the variation in the total column density of the MOT as the repump detuning changes, but a direct comparison between the experimental and theoretical populations can be made by looking at the fractional population difference. Figure \ref{fig:rep_pops}(b) shows that the normalized difference continues to increase with repump detuning. The theoretical populations obtained from the master equation reproduce this behavior. 

\section{\label{sec:conc} Conclusion}
We have studied optical gain in an operating magneto-optical trap of $^{39}$K atoms. The gain was shown to arise from Raman transitions between the hyperfine ground states, which has not been observed in-situ with more commonly used species such as rubidium or cesium. The conditions for such gain are set by the relatively small excited-state hyperfine splittings. First, the cooling light is closer to resonance with an open transition than the cycling transition, leading to rapid optical pumping into the lower ground state. This in turn necessitates a relatively large intensity for the repump light, which must also be red-detuned to reduce heating. Even with these adjustments, we typically have a larger population in the lower ground state, creating the effective inversion needed for gain. For completeness, we note that hyperfine Raman gain also has been exploited to impressive effect in a wide range of more complex experiments with cold rubidium or cesium \cite{McKeever2003, Schneble2004, Yoshikawa2004, Yoshikawa2005, Vrijsen2011, Bohnet2012, Baudouin2013}. Our demonstration of in-situ Raman gain in a potassium MOT suggests a comparatively simple and robust experimental platform for studying steady-state gain and lasing with cold atoms.
 
Our measurements of the potassium ring laser have shown that the gain is effectively homogeneously broadened, despite the presence of inhomogeneous magnetic fields and Doppler shifts \cite{Megyeri2018}. This prevents simultaneous bidirectional lasing into the counter-propagating modes of the ring cavity in favor of random switching between modes. By intermittently shutting off the MOT fields and applying a separate far-detuned Raman pump, it should be possible to realize homogeneous gain linewidths on the order of only a few Hz \cite{Bohnet2012}. In this case inhomogeneous Doppler broadening should enable simultaneous lasing \cite{SargentMurray1974}.

\acknowledgments{This work was funded by the UK Engineering and Physical Sciences Research Council (EP/J016985/1). We are grateful to the members of the Quantum Matter research group at the University of Birmingham for useful discussions and the loan of equipment.}

\bibliographystyle{unsrt}
\bibliography{paper}% Produces the bibliography via BibTeX.

\end{document}